\documentstyle[preprint,tighten,aps,psfig]{revtex} %preprints

\begin{document}

\draft
\title{Kinetic Energy Decay Rates of Supersonic and Super-Alfv\'enic
Turbulence in Star-Forming Clouds}

\author{Mordecai-Mark Mac Low\cite{mpi}, Ralf S. Klessen\cite{mpi}, Andreas Burkert\cite{mpi}}

\address{Max-Planck-Institut f\"ur Astronomie, K\"onigstuhl 17,
D-69117 Heidelberg, Germany}

\author{Michael D. Smith\cite{mds}}
\address{Astronomisches Institut der Universit\"at W\"urzburg, Am Hubland,
D-97074 W\"urzburg, Germany}

\date{Submitted to {\em Phys. Rev. Letters} 28 November 1997}
\maketitle
\begin{abstract}
We present numerical studies of compressible, decaying turbulence,
with and without magnetic fields, with initial rms Alfv\'en and Mach
numbers ranging up to five, and apply the results to the question of
the support of star-forming interstellar clouds of molecular gas.  We
find that, in 1D, magnetized turbulence actually decays faster than
unmagnetized turbulence.  In all the regimes that we have studied 3D
turbulence---super-Alfv\'enic, supersonic, sub-Alfv\'enic, and
subsonic---the kinetic energy decays as $(t-t_{\rm 0})^{-\eta}$, with
$0.85 < \eta < 1.2$.  We compared results from two entirely different
algorithms in the unmagnetized case, and have performed extensive
resolution studies in all cases, reaching resolutions of $256^3$ zones
or 350,000 particles.  We conclude that the observed long lifetimes
and supersonic motions in molecular clouds must be due to external
driving, as undriven turbulence decays far too fast to explain the
observations.
\end{abstract}

%Turbulence simulation and modeling, Isotropic/homogeneous
%turbulence, High Reynolds number turbulence, Supersonic and
%hypersonic flows, MHD and electrohydrodynamics, Astrophysical MHD &
%plasmas, Astrophysical hydrodynamics, Molecular clouds (galactic), Star
%Formation, Physical properties of the ISM (galactic), Computational
%methods in fluid dynamics, Finite-difference methods

\narrowtext

Star-forming clouds of interstellar gas emit strongly in molecular
emission lines; temperatures derived from these lines show that the
linewidths greatly exceed the thermal sound speed $c_{\rm s}$ in these
clouds.  With densities of order $n \sim 10^3 - 10^5$~protons per
cm$^{3}$, the gas in these clouds can be described by an isothermal
equation of state due to the efficient radiative cooling allowed by
the many low-lying molecular transitions \cite{molrev}.  Cloud
lifetimes are of order $3 \times 10^7$ yr \cite{bs80}, while free-fall
gravitational collapse times are only $t_{\rm ff} = (1.4 \times 10^6
\mbox{ yr})(n/10^3 \mbox{ cm}^{-3})^{-1/2}$. In the absence of
non-thermal support, these clouds should collapse and form stars in a
small fraction of their observed lifetime.  Supersonic hydrodynamical
(HD) turbulence is suggested as a support mechanism by the observed
broad lines, but was dismissed because it would decay in times of
order $t_{\rm ff}$.  A popular alternative has been sub- or
trans-Alfv\'enic magnetohydrodynamical (MHD) turbulence, which was
thought to decay an order of magnitude more slowly \cite{am75}.
However, analytic estimates and computational models suggest that {\em
incompressible} MHD turbulence decays \cite{bis94,hos95,pol95,gpp97}
as $(t-t_{\rm 0})^{-\eta}$, with a decay rate $2/3 < \eta < 1.0$,
where $t_{\rm 0}$ is some characteristic time scale, while
incompressible HD turbulence has been experimentally measured
\cite{exp,smith93} to decay with $1.2 < \eta < 2$.  The difference in
decay rates between incompressible HD and MHD turbulence is clearly
not as large as had been suggested for compressible astrophysical
turbulence.  In this paper we use high-resolution, three-dimensional
(3D) simulations to compute the decay rates of compressible,
homogeneous, isothermal, decaying turbulence with supersonic,
sub-Alf\'enic, and super-Alfv\'enic root-mean-square (rms) initial
velocities $v_{\rm rms}$, and show that the decay rates in these
physical regimes, $0.85 < \eta < 1.2$, strongly resemble the
incompressible results.

\paragraph{Previous Work}

The decay of the kinetic energy $E_K$ of incompressible HD turbulence
has been explored in some detail. The energy is predominantly held
within low wavenumber modes. Dissipation, on the other hand, occurs
predominantly from the high wavenumber modes.  Therefore, the decay
rate does not generally depend on the details of the dissipation
process.  Rather, it is controlled by the efficiency of energy
transfer from the low to high wavenumber modes due to vortex
interactions, nonlinear wave interactions, or other processes.  This
leads either to the Kolmogorov decay rate $\eta = 10/7$ for the
kinetic energy, accompanied by a growth in the external scale of the
turbulence ${\cal L} \propto t^{2/7}$ \cite{k41}, or to the closure
model law $\eta = 1 + (s-1)/(s+3)$ and ${\cal L} \propto t^{2/(s+3)}$,
depending on the injected energy spectrum in the low wavenumber region
($P(k) \propto k^s$) and the spatial dimension D \cite{les97}.  Hence
values for the decay rate $\eta$ as low as unity, are predicted, much
lower than Kolmogorov.  However, these results are from studies of
spatially free turbulence. When turbulence is confined \cite{smith93},
a much higher rate is found experimentally: $\eta \rightarrow 2$.
These confined turbulence experiments have been modeled with a mean
field theory \cite{loh94}.

The decay behaviour of incompressible MHD turbulence, in contrast, is
controversial.  A two-dimensional (2D) analysis with constant mean
square magnetic potential yields $\eta =1$, a result indeed backed up
by numerical studies in 2D \cite{bw89}.  A dimensional analysis in 3D
assuming magnetic helicity invariance \cite{h84} yields $\eta =2/3$,
although low-resolution numerical results \cite{hos95,gpp97} suggest
$\eta \sim 1$.

Compressibility introduces an alternative type of complexity \cite{lele94}.
Nevertheless, the 3D decay problem has been modelled numerically
\cite{por94}.  Although the evolution of the spectral development and
spatial structures was explored there, the decay rate was only treated
in passing, and definitive results are hard to derive. 

Unlike terrestrial turbulence, astrophysical turbulence usually
involves full MHD compressible flow.  Numerical models of
one-dimensional (1D), isothermal, compressible, strongly magnetized,
decaying and forced turbulence have been performed by \cite{go96}, who
found decay rates rather lower than those discussed above.  A recent
3D study \cite{pn97} broke new ground by examining both weakly and
strongly magnetized isothermal, compressible turbulence; however the
rather low energy decay rates shown there were almost entirely
dependent on the initial conditions chosen, as explained in
\cite{mm97}.

\paragraph{Numerical Techniques}
For our HD models of strongly supersonic turbulence we use two
entirely different HD methods---a second-order, Eulerian,
finite-difference code, and a smoothed particle hydrodynamics (SPH)
code---while for our MHD models we use only the finite-difference code.

This finite-difference code is the well-tested MHD code ZEUS
\cite{sn92}, which uses second-order \cite{vl77} advection, and a
consistent transport algorithm for the magnetic fields \cite{ct}.  It
resolves shocks using a standard von Neumann artificial viscosity, but
otherwise includes no explicit viscosity, relying on numerical
viscosity to provide dissipation at small scales.  This should
certainly be a reasonable approximation for shock-dominated flows, as
most dissipation occurs in the shock fronts, where the artificial
viscosity dominates in any case.  The relative simplicity of this
Eulerian formulation allows us to perform resolution studies showing
that our major results are, in fact, independent of the resolution,
and thus of the strength of numerical viscosity.

SPH is a particle based approach to solving the HD equations
\cite{sph}, in which the system is represented by an ensemble of
particles, each carrying mass, momentum, and fluid properties such as
pressure, temperature, and internal energy special-purpose processor
GRAPE to accelerate computation of nearest-neighbor lists
\cite{grape}, allowing us to use as many as 350,000 particles.

\paragraph{Initial Conditions}
We chose initial conditions for our models inspired by the popular
idea that setting up velocity perturbations with an initial power
spectrum $P(k) \propto k^\alpha$ in Fourier space similar to that of
developed turbulence would be in some way equivalent to starting with
developed turbulence \cite{pn97,por94}.  Observing the development of
our models, it became clear to us that, especially in the supersonic
regime, the loss of phase information in the power spectrum allows
extremely different gas distributions to have the same power spectrum.
For example, supersonic, HD turbulence has been found in simulations
\cite{por94} to have a power spectrum $\alpha = -2$.  However, any
single, discontinuous shock wave will also have such a power spectrum,
as that is simply the Fourier transform of a step function, and taking
the Fourier transform of many shocks will not change this power law.
Nevertheless, most distributions with $\alpha = -2$ do not contain shocks.

After experimentation, we decided that the quickest way to generate
fully developed turbulence was with perturbations having a flat power
spectrum $\alpha = 0$ with a cutoff at moderate wavenumber (typically
$k_{\rm max} = 8$).  We set up velocity perturbutions drawn from a
Gaussian random field fully determined by its power spectrum in
Fourier space following the standard procedure: for each wavenumber
${\bf k}$ we randomly select an amplitude from a Gaussian distribution
centered on zero and with width $P(k)= P_{\rm 0} k^\alpha$ with
$k=|{\bf k}|$, and a phase between zero and $2\pi$.  We then transform
the field back into real space to obtain the velocity in each zone.
This is done independently for each velocity component.  For the SPH
calculation the velocities defined on the grid are assigned onto
individual particles using the ``cloud-in-cell'' scheme \cite{he88}.
In all of our models we take $c_{\rm s} = 0.1$, initial density
$\rho_{\rm 0} = 1$, and we use a periodic grid with sides $L = 2$
centered on the origin.  These parameter choices define our unit
system.

\begin{figure}
\psfig{file=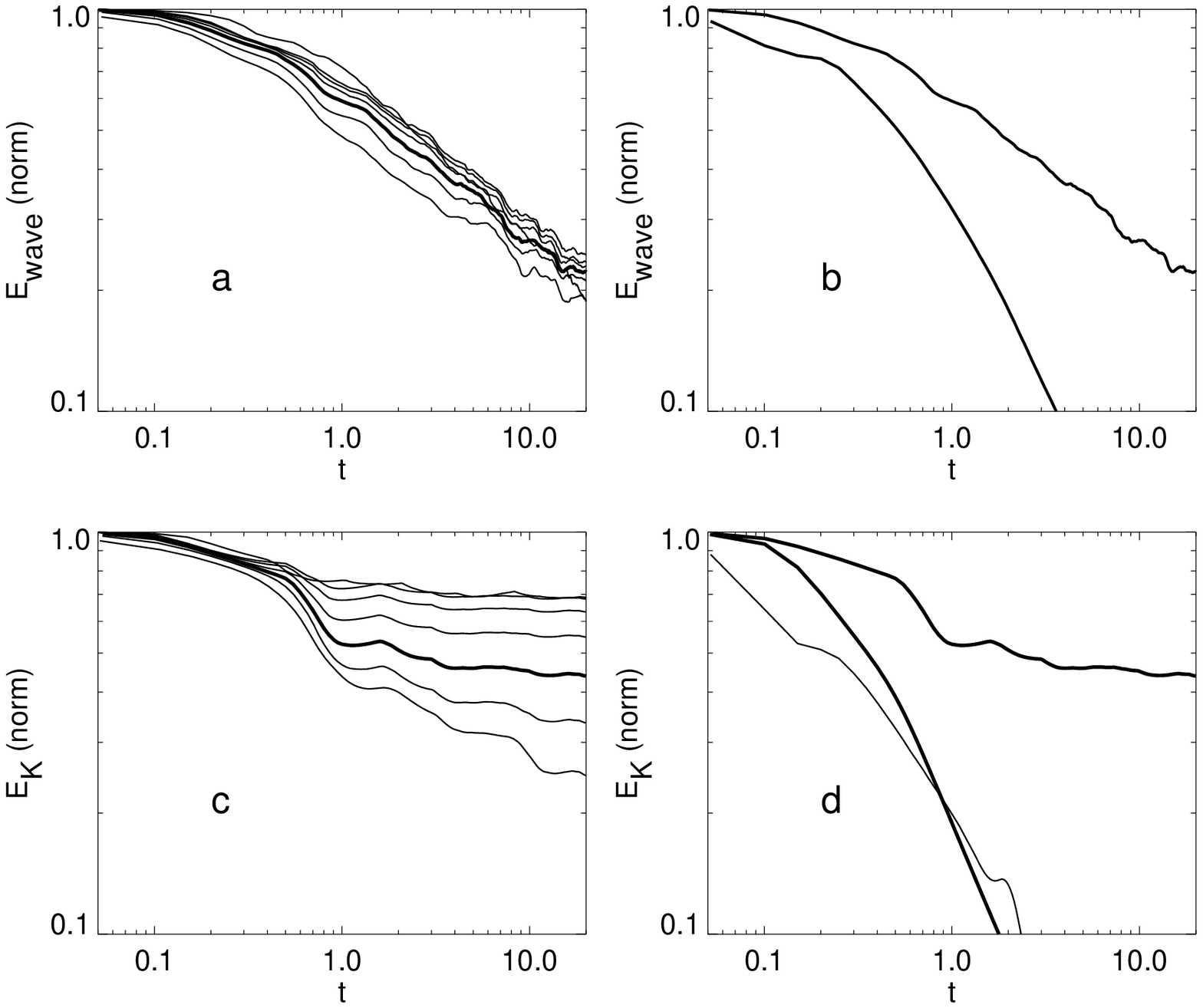,angle=0,width=\hsize,clip=}
\caption{Isothermal, $M=5$ models with ZEUS.  {\em a)} Wave energy
decay of MHD models with $A=1$ and resolutions ranging from 32 to 4096
zones; the 256 zone model is highlighted. {\em b)} Comparison of the
same 256 zone 1D MHD model ({\em upper line}) to $256^3$ zone 3D MHD
Model Q ({\em lower line}). {\em c)} Kinetic energy decay of HD models
with resolutions ranging from 32 ({\em lowest}) to 4096 ({\em
highest}) zones; the 256 zone model is highlighted. {\em d)}
Comparison of the same 256 zone 1D HD model ({\em upper thick line})
to $256^3$ zone HD Model D ({\em lower thick line}) , and to $256^3$
zone MHD Model Q ({\em thin line})}
\label{1D}
\end{figure}

\paragraph{One-Dimensional Results}
To verify our numerical methods, we reproduced the 1D, MHD
results of Gammie \& Ostriker \cite{go96}.  Fig.~\ref{1D}{\em a} shows
the results of a resolution study comparable to their Figure~1, with
$M = 5$, initial uniform field parallel to the $x$-axis, and
initial rms Alfv\'en number $A = v_{\rm rms}/v_{\rm A} = 1$, where
$v_{\rm A}^2 = B^2 / 4\pi \rho_{\rm 0}$.  Note that $t=20$ in our
units corresponds to $t=1$ in theirs.  Aside from a rather
faster convergence rate in our study, attributable to the details of
our choice of initial conditions, we reproduce excellently their
result: a decrease in wave energy $E_{\rm wave} = E_{\rm K} + (B_{\rm
y}^2 + B_{\rm z}^2)/8\pi$ by a factor of five in one sound-crossing
time $L/c_{\rm s}$.

We then extended our study by examining the equivalent HD problem, as
shown in Fig.~\ref{1D}{\em c}, only to find that the decay rate of HD
turbulence in 1D is significantly {\em slower} than that of MHD
turbulence.  This appears to be due to the sweeping up of slower
shocks by faster ones in the HD case, resulting in pure Bergers
turbulence, with linear velocity profiles between widely separated
shocks exactly as predicted \cite{berg}.  The result is that there are
very few dissipative regions, and energy is only lost very slowly.  In
contrast, multiple wave interactions occur in the MHD case producing
many dissipative regions and so faster dissipation.

Finally we compared 1D models with 256 zone resolution to equivalent
3D models with $256^3$ zones.  The 3D model loses energy far faster
than the 1D model in both the HD case shown in the thick lines in
Fig.~\ref{1D}{\em d} and the MHD case shown in Fig.~\ref{1D}{\em b}.
The increased number of degrees of freedom available in 3D produces
more shocks and interaction regions, resulting in increased energy
dissipation.

\begin{figure}
\psfig{file=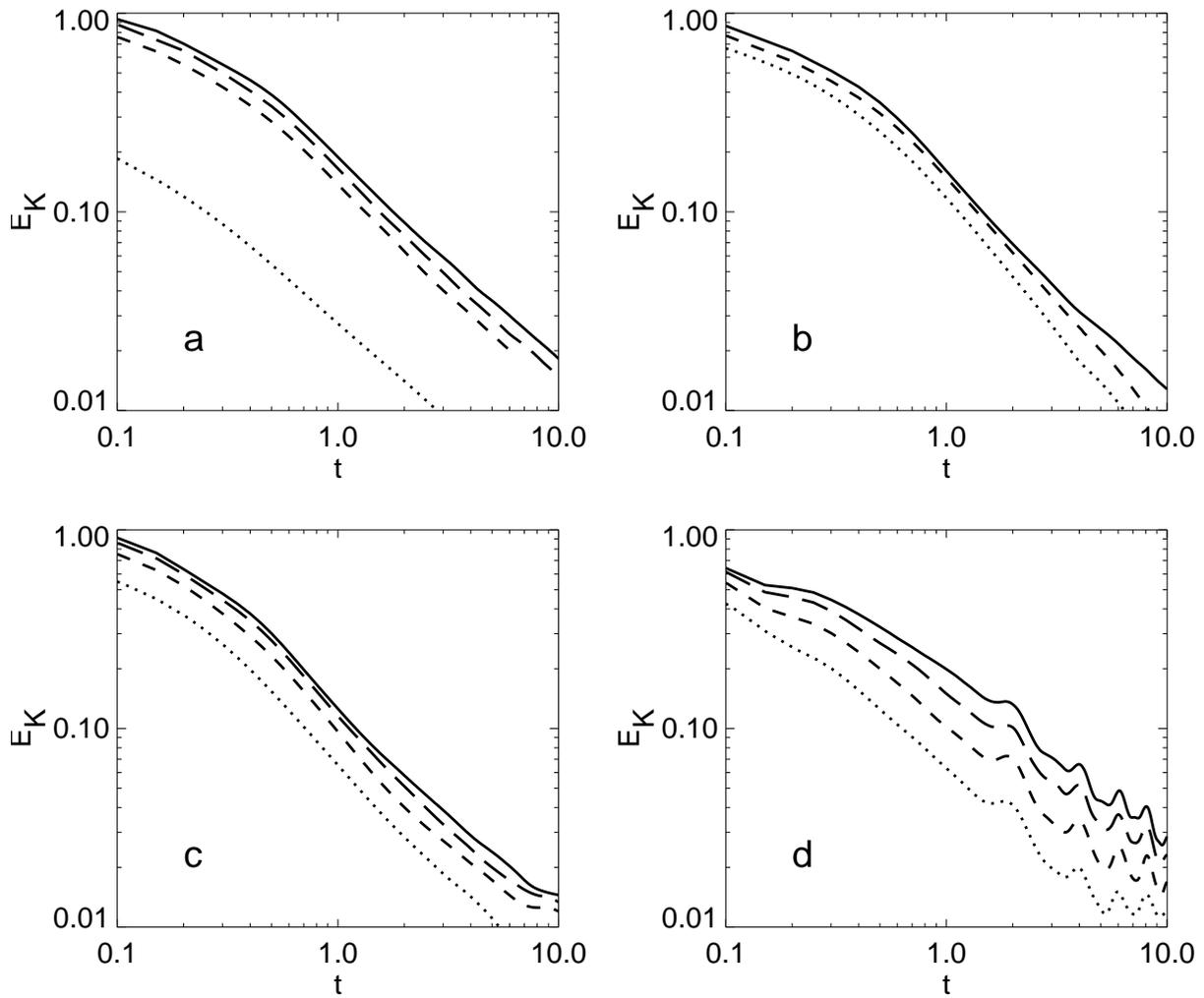,angle=0,width=\hsize,clip=}
\caption{3D resolution studies for M=5, isothermal
models.  ZEUS models have $32^3$ ({\em dotted}), $64^3$ ({\em short
dashed}), $128^3$ ({\em long dashed}), or $256^3$ ({\em solid}) zones,
while the SPH models have 7000 ({\em dotted}), 50,000 ({\em short
dashed}), or 350,000 ({\em solid}) particles. Panels show
{\em a)} HD runs with ZEUS, {\em b)} HD runs
with SPH, {\em c)} $A=5$ MHD runs with ZEUS, and {\em d)} $A=1$ MHD
runs with ZEUS. }
\label{3D}
\end{figure}

\paragraph{Three-Dimensional Results}
We next performed resolution studies using ZEUS for three different
cases with no field, weak field and strong field as described in
Fig.~\ref{3D}, and summarized in Table~\ref{runs}.  The weak field
models have an initial ratio of thermal to magnetic pressure $\beta =
2$, while the strong field models have $\beta = 0.08$.  We ran the
same HD model with the SPH code to demonstrate that our results are
truly independent of the details of the viscous dissipation, and so
that our lack of an explicit viscosity does not affect our results.
We also ran two models (R, S) with adiabatic index $\gamma = 1.4$, and
an isothermal model (T) with initial $M = 0.1$ to provide a point of
direct comparison between our results and those for incompressible,
Navier-Stokes turbulence.

The kinetic energy decay curves for the four resolution studies are
shown in Fig.~\ref{3D}.  For each of our runs we performed a
least-squares fit to the power-law portion of the kinetic energy decay
curves, and report the corresponding decay rate $\eta$ in
Table~\ref{runs}.  These results appear converged at the 5--10\%
level; it is very reassuring that the different numerical methods
converge to the same result for the HD case.

We find that highly compressible, isothermal turbulence (Model D)
decays somewhat more slowly, with $\eta = 0.98$, than less
compressible, adiabatic turbulence (Model R), with $\eta = 1.2$, or
than incompressible turbulence (Model T), with $\eta = 1.1$ (also see
\cite{smith93,loh94}).  Adding magnetic fields (Models L, Q, and S)
decreases the decay rate somewhat further in the isothermal case to
$\eta \sim 0.9$, with very slight dependence on the field strength or
adiabatic index.

\paragraph{Conclusions}

We can draw conclusions for turbulence theory from our models that
have significant astrophysical implications. What we find remarkable
is how closely our results resemble the incompressible results,
despite the difference in dissipation mechanisms.  In incompressible
hydrodynamics, kinetic energy is dissipated in vortexes at the
smallest scales while in supersonic compressible turbulence, kinetic
energy is dissipated in shock waves, and in MHD turbulence kinetic
energy is dissipated in the interactions of nonlinear Alfv\'en waves,
yet the resulting decay rates differ only slightly.  The difference
between our 1D and 3D HD results suggests that somehow the space
density of dissipative regions is the determining factor, and that in
3D it is somehow quite independent of the detailed physics.

The clear astrophysical implication of these models is that even
strong magnetic fields, with the field in equipartition with the
kinetic energy, cannot prevent the decay of turbulent motions on
dynamical timescales far shorter than the observed lifetimes of
molecular clouds \cite{zj83}.  The significant kinetic energy observed
in molecular cloud gas must be supplied more or less continuously.  If
turbulence supports molecular clouds against star formation, it must
be constantly driven, by stellar outflowsw \cite{sn80}, photoionization
\cite{photo}, galactic shear \cite{f81}, or some combination of these
or other sources.

%\acknowledgments 
Some computations presented here were performed at the Rechenzentrum
Garching of the MPG.  ZEUS was used courtesy of the Laboratory for
Computational Astrophysics at the NCSA.  MDS thanks the DFG for
financial support.

% tables follow here
%
\begin{table}
\caption{Power law of kinetic energy decay $\eta$ (with formal errors
from the least squares fits) for the 3D models discussed.  Initial rms
Mach number $M$ and Alfv\'en number $A$, and adiabatic index $\gamma$
are given, along with the resolution (in zones per side or thousands
of particles) and code used.  Boldface indicates highest resolution.}
\label{runs}
\begin{tabular}{lllllll}
Mod & code & res & $\gamma$ & M & A  & $\eta$ \\
A   & ZEUS & 32  & 1 & 5 & $\infty$ & 0.93 $\pm$ 0.001 \\
B   & ZEUS & 64  & 1 & 5 & $\infty$ & 1.1 $\pm$ 0.003 \\
C   & ZEUS & 128 & 1 & 5 & $\infty$ & 1.0 $\pm$ 0.002 \\
\bf D & \bf ZEUS & \bf 256 & \bf 1 & \bf 5 & \boldmath $\infty$ & \boldmath \bf
    0.98$\pm$ 0.001  \\
E   & SPH  & 7   & 1 & 5 & $\infty$ & 1.3 $\pm$ 0.005 \\
F   & SPH  & 50  & 1 & 5 & $\infty$ & 1.2 $\pm$ 0.001 \\
\bf G & \bf SPH  & \bf 350 & \bf 1 & \bf 5 & \boldmath $\infty$  &
    \boldmath \bf 1.1 $\pm$ 0.004  \\
H   & ZEUS & 32  & 1 & 5 & 5 & 0.89$\pm$ 0.02  \\
J   & ZEUS & 64  & 1 & 5 & 5 & 0.80$\pm$ 0.01  \\
K   & ZEUS & 128 & 1 & 5 & 5 & 0.86$\pm$ 0.01  \\
\bf L & \bf ZEUS & \bf 256 & \bf 1 & \bf 5 & \bf 5 & \boldmath \bf 
    0.91$\pm$ 0.006  \\
M   & ZEUS & 32  & 1 & 5 & 1 & 0.67$\pm$ 0.02  \\
N   & ZEUS & 64  & 1 & 5 & 1 & 0.80$\pm$ 0.02  \\
P   & ZEUS & 128 & 1 & 5 & 1 & 0.83$\pm$ 0.02  \\
\bf Q & \bf ZEUS & \bf 256 & \bf 1 & \bf 5 & \bf 1 & \boldmath \bf
    0.87$\pm$ 0.02  \\
\bf R & \bf ZEUS & \bf 256 & \bf 1.4 & \bf 5 & \boldmath $\infty$ &
    \boldmath \bf 1.2 $\pm$ 0.006 \\
\bf S & \bf ZEUS & \bf 256 & \bf 1.4 & \bf 5 & \bf 1 & \boldmath \bf
    0.94$\pm$ 0.009  \\
\bf T & \bf ZEUS & \bf 256 & \bf 1 & \bf 0.1 & \boldmath $\infty$ &
    \boldmath \bf 1.1 $\pm$ 0.007  \\
\end{tabular}
\end{table}


\begin{references}

\bibitem[*]{mpi} Email addresses: mordecai, burkert, klessen@mpia-hd.mpg.de

\bibitem[\dagger]{mds} Email address: smith@astro.uni-wuerzburg.de

\bibitem{molrev} For an overview of molecular cloud physics and
observations, see reviews by B. G. Elmegreen and by L. Blitz
in {\em Protostars and Planets III}, edited by E. H. Levy and
J. I. Lunine (University of Arizona Press, Tucson, 1993), p. 97 and
125, respectively

\bibitem{bs80} L. Blitz and F. H. Shu, Astrophys. J. {\bf 238}, 148
(1980)

\bibitem{am75} M. Arons and C. Max, Astrophys. J. {\bf 196}, L77 (1975).
Also see \cite{go96} and references therein. 

\bibitem{bis94} D. Biskamp, {\em Nonlinear Magnetohydrodynamics}
(Cambridge University Press, Cambridge, England, 1994).

\bibitem{hos95} M. Hossain, P. Gray, D. Pontius, W. Matthaeus, and
S. Oughton, Phys. Fluids {\bf 7}, 2886 (1995).

\bibitem{pol95} H. Politano, A. Pouquet, and P. L. Sulem in {\em
Small-Scale Structures in Fluids and MHD}, edited by M. Meneguzzi,
A. Pouquet, and P. L. Sulem, Springer-Verlag Lecture Notes in Physics
Vol. 462 (Springer-Verlag, Berlin, 1995), p. 281.

\bibitem{gpp97} S. Galtier, H. Politano, and A. Pouquet,
Phys. Rev. Lett. {\bf 79}, 2807 (1997).

\bibitem{exp} G. Comte-Bellot and S. Corrsin, J. Fluid Mech. {\bf 25},
657 (1966); Z. Warhaft and J. Lumley, J. Fluid Mech. {\bf 88}, 659
(1978).

\bibitem{smith93} M. R. Smith, R. J. Donnelly, N. Goldenfeld, and
W. F. Vinen, Phys. Rev. Lett. {\bf 71}, 2583 (1993).

\bibitem{k41} A. N. Kolmogorov, Dokl. Acad. Nauk SSSR {\bf 30}, 9
(1941); C. F. von Weizs\"acker, Z. Phys. {\bf 124}, 614 (1948);
L. D. Landau and E. M. Lifshitz, {\em Fluid Mechanics} (Pergamon
Press, Oxford, 1959), p. 143.

\bibitem{les97} M. Lesieur, {\em Turbulence in Fluids} (Kluwer
Academic Publishers, Dordrecht, 1997).

\bibitem{loh94} D. Lohse, Phys. Rev. Lett. {\bf 73}, 3223 (1994).

\bibitem{bw89} D. Biskamp and H. Welter, Phys. Fluids B {\bf 1}, 1964
(1989).
 
\bibitem{h84} T. Hatori, J. Phys. Soc. Japan {\bf 53}, 2539 (1984).

\bibitem{lele94} S. K. Lele, Ann. Rev. Fluid Mech. {\bf 26}, 211
(1994).

\bibitem{por94} D. H. Porter, A. Pouquet, and P. R. Woodward,
Phys. Rev. Lett. {\bf 68}, 3156 (1992);  D. H. Porter, A. Pouquet, and
P. R. Woodward, Phys. Fluids {\bf 6}, 2133 (1994).

\bibitem{go96} C. F. Gammie and E. C. Ostriker, Astrophys. J. {\bf
466}, 814 (1996).

\bibitem{pn97} P. Padoan, and \AA. Nordlund, Astrophys. J., submitted
(1997), astro-ph/9703110.

\bibitem{mm97} M.-M. Mac Low in press in {\em The Orion Nebula
Revisited}, edited by M. J. McCaughrean and A. Burkert (San Francisco,
Publ. Astron. Soc. Pacific, 1998), astro-ph/9711349.

\bibitem{sn92} J. M. Stone and M. L. Norman, Astroph. J. {\bf 80},
753 (1992);  J. M. Stone and M. L. Norman, Astroph. J. {\bf 80}, 791
(1992).

\bibitem{vl77} B. van Leer, J. Comput. Phys. {\bf 23}, 276  (1977).

\bibitem{ct} C. Evans, and J. F. Hawley, Astrophys. J. {\bf 33}, 659 (1988).

\bibitem{sph} Reviews of the SPH method are given by W. Benz
in {\em The Numerical Modelling of Nonlinear Stellar Pulsations},
edited by J. R. Buchler (Kluwer, Dordrecht, Netherlands, 1990),
p. 269, and by J. J. Monaghan, Ann. Rev. Astron. Astroph. {\bf
30}, 543 (1992).

\bibitem{grape} The GRAPE project is described by
T. Ebisuzaki, J. Makino, T. Fukushige, M. Taiji, D. Sugimoto, T. Ito,
and S. Okumura, Publ. Astron. Soc. Japan {\bf 45}, 269 (1993); and the
application to SPH is described in M. Steinmetz,
M. N. Roy. Astron. Soc. {\bf 278}, 1005 (1996). 

\bibitem{he88}
R. W. Hockney and J.~W. Eastwood, {\em Computer Simulation Using
Particles} (Institute of Physics, Bristol, England, 1988) 

\bibitem{berg} See p. 239 of \cite{les97}.

\bibitem{zj83} This was implied, though not emphasized, by the
analytic study of E. G. Zweibel, and K. Josafattson,
Astrophys. J. {\bf 270}, 511 (1983), as well as by the numerical study
of \cite{pn97}, though with the problems discussed in paragraph {\em a}.

\bibitem{sn80} e.\ g.\ J. Silk and C. Norman, in {\em Interstellar Molecules,
IAU Symposium 87}, edited by B. H. Andrew (Reidel, Dordrecht, 1980),
p. 165.

\bibitem{photo} C. F. McKee, Astrophys. J. {\bf 345}, 782 (1989);
F. Bertoldi and C. F. McKee, in {\em Amazing Light},
edited by R. Y. Chiao (Springer, New York, 1996), p. 41.

\bibitem{f81} R. C. Fleck, Jr., Astrophys. J. {\bf 246}, L151 (1981).

\end{references}
\end{document}